\documentclass[12pt]{article}

\def\[{\left\lbrack}
\def\]{\right\rbrack}

\def\({\left(}
\def\){\right)}
\def\ih{\'\i}

\renewcommand{\baselinestretch}{1.5} 
  
\usepackage{graphicx}

\label{}
\begin{document}
\date{}
\title{Fundamental Constants, Entropic Gravity and Nonextensive Equipartition Theorem}
\author{Jorge Ananias Neto\footnote{e-mail: jorge@fisica.ufjf.br}\\
Departamento de F\ih sica, ICE, \\ Universidade Federal de Juiz de Fora, 36036-900,\\ Juiz de Fora, MG, Brazil }

\maketitle

\begin{abstract}
By using the Verlinde's formalism, we propose that the positive numerical factor, in which  Klinkhamer states that it is necessary to define the fundamental length, can be  associated to the parameter $q$ of the Tsallis' nonextensive statistical mechanics.

\end{abstract}


\newpage

\section{Introduction}
The Newton constant G arises from the three fundamental constants
\begin{eqnarray}
\label{gc}
G = \frac{l_p^2\, c^3}{\hbar},
\end{eqnarray}
where $l_p$ is the Planck length, $c$ is the velocity of light and $\hbar$ is the Planck constant. The Planck constant describes the role of the quantum matter while the Planck length probably governs the properties of the unknown quantum gravity. Klinkhamer\cite{Klink} has suggested that the Planck length can be expressed in terms of more fundamental constants

\begin{eqnarray}
\label{lp}
l_p^2 = f\, l^2, 
\end{eqnarray}
where $f$ is a positive numerical factor associated, in principle, to the microscopic degrees of freedom on the holographic screen  and the constant $l$, in which it  has the dimension of length, regulates the quantum space time. According to the Klinkhamer point of view, this fundamental length should be independent of the direct presence of matter or nongravitational fields. Moreover, one believes that relation (\ref{lp}) must be universal, i.e., it is, in principle, valid for any physical system. Therefore, nature should have three fundamental independent constants that are $\hbar$, $c$ and $l$. In the same line, Sahlmann\cite{Sa} has proposed two different microscopic structures for the holographic screen where he has obtained specific values for $f$ and $l$. Shao and Ma\cite{SM} have also suggested that the Newtonian constant $G$ does not play a fundamental role any more and a new fundamental length must be introduced where its value can be measure by Lorentz violation experiments.

In this work, we apply a nonextensive statistical mechanics formalism\cite{Tsa} in the Verlinde's framework of gravitational theory\cite{Ver}. We will see that it is possible to associate the positive numerical factor $f$  with the parameter $q$ of the Tsallis' nonextensive thermostatistics and, consequently, to disentangle $f$ and $l$ from the product, Eq.(\ref{lp}).
 This approach is an extension of the standard Boltzmann-Gibbs statistical mechanics and has been successfully applied in many different physical systems. Among them we can mention Levy-type anomalous diffusion\cite{Levy}, turbulence in a pure-electron plasma
\cite{CT} and gravitational systems\cite{Ng}. It is worth mentioning that the articles of the last reference state that the gravitational interactions (i.e., systems endowed with long range interactions) lead to a nonextensive statistical mechanics.

\section{Brief Review of Verlinde's Formalism}
\label{bfv}

The connection between the Newton constant $G$ and the fundamental constants can be made by a formalism proposed
by E. Verlinde in which the gravitational acceleration is obtained by using, basically, the holographic principle and the equipartition law of energy. This is an important result and,  at moment, many authors have working actively
on this subject\cite{sev}. It is opportune to comment here that  there is, at moment, an interesting debate about the validity of a particular entropy's formula proposed by Verlinde in the context of the neutron quantum states in the Earth's gravitational field\cite{Ko, Ch}. However,  in our approach we use a well accepted procedure which was initially developed by T.  Padmanabhan\cite{Pa}. The Verlinde's proposal considers a spherical surface as the holographic screen, with a particle of mass M positioned in its center. A holographic screen can be thought as a storage device for information. The number of bits (The term bit signifies the smallest unit of information in the holographic screen) is assumed to be proportional to the area $A$ of the holographic screen

\begin{eqnarray}
\label{bits}
N = \frac{A }{l_p^2},
\end{eqnarray}
where $ A = 4 \pi r^2 $ and $l_p = \sqrt{\frac{G\hbar}{c^3}}$ is the Planck length. In the Verlinde's formalism we assume that the total energy of the bits on the screen is given by the equipartition law of energy

\begin{eqnarray}
\label{eq}
E = \frac{1}{2}\,N k_B T.
\end{eqnarray}
It is important to comment here that the traditional equipartition theorem, Eq.(\ref{eq}), is derived from the usual Boltzmann-Gibbs thermostatistics. In a nonextensive thermostatistics scenario, the equipartition law of energy will be modified in a sense that a nonextensive parameter $q$ will be introduced in its expression.
Considering that the energy of the particle inside the holographic screen is equally divided on all bits then we can write the equation

\begin{eqnarray}
\label{meq}
M c^2 = \frac{1}{2}\,N k_B T.
\end{eqnarray}
Using Eq.(\ref{bits}), and the Unruh temperature formula\cite{Unr}

\begin{eqnarray}
\label{un}
k_B T = \frac{1}{2\pi}\, \frac{\hbar a}{c},
\end{eqnarray}
we are  in a position to derive the famous (absolute) gravitational acceleration formula

\begin{eqnarray}
\label{acc}
a =  \frac{l_p^2 c^3}{\hbar} \, \frac{ M}{r^2}\nonumber\\ 
= G \, \frac{ M}{r^2}.
\end{eqnarray}
We can observe that from Eq.(\ref{acc}) the Newton constant $G$ is just written in terms of the fundamental constants, Eq.(\ref{gc}). Using Eq.(\ref{lp}), we can write the Newton constant as

\begin{eqnarray}
G = \frac{f l^2 c^3}{\hbar}.
\end{eqnarray}
The Planck length $l_p = 1.6162 \times 10^{-35}\, m$ (and consequently the product $l_p^2 = f\,l^2\,$) is fixed by a precise value of $G$ measured by experimental techniques that is, in a concise form, $ G = 6.6743(7) \times 10^{-11}m^3 kg^{-1} s^{-2}$\cite{Codata}.

\section{The Nonextensive Equipartition Theorem and Its Application in the Verlinde's Formalism}

An important formulation of a nonextensive  Boltzmann-Gibbs thermostatistics has been proposed by Tsallis in which the entropy is given by the formula

\begin{eqnarray}
\label{nes}
S_q =  k_B \, \frac{1 - \sum_{i=1}^W p_i^q}{q-1}\;\;\;\;\;\; (\sum_{i=1}^W p_i = 1),
\end{eqnarray}
where $p_i$ is the probability of the system to be in a microstate, $W$ is the total number of configurations and $q$ is a real parameter quantifying the degree of nonextensivity. 
The definition of entropy (\ref{nes}) has as motivation multifractals systems and possesses the usual properties of positivity, equiprobability, concavity and irreversibility.
It is important to note that Tsallis' formalism contains the Boltzmann-Gibbs statistics as a particular case in the limit $ q \rightarrow 1$ where the usual additivity of entropy is recovered. Plastino and Lima\cite{PL} by using a generalized velocity distribution for free particles\cite{SPL}

\begin{eqnarray}
f_0(v) = B_q \[ 1-(1-q) \frac{m v^2}{2 k_B T} \]^{1/1-q},
\end{eqnarray}
where $B_q$ is a $q$-dependent normalization constant, $m$ and $v$ is a mass and velocity of the particle, respectively, have derived a nonextensive equipartition law of energy whose expression is given by

\begin{eqnarray}
\label{ge}
E = \frac{1}{5 - 3 q} N k_B T,
\end{eqnarray}
where the range of $q$ is $ 0 \le q < 5/3 $.  For $ q=5/3$ (critical value) the expression of the equipartition law of energy, Eq.(\ref{ge}), diverges. It is easy to observe that for $ q = 1$,  the classical equipartition theorem for each microscopic degrees of freedom is retrieved. It is important to mention that the virial theorem is not modified in this nonextensive  thermostatistics formalism\cite{MPP}.

To investigate the $q$-dependence of the numerical factor $f$, we consider the generalized equipartition theorem, Eq.(\ref{ge}), within the context of the Verlinde's formalism,  i.e. the bits now obey a nonextensive statistical mechanics. The Verlinde's formalism is notable since it allows to infer changes in the physical properties of a particular gravitational system when the equipartition law of energy is modified under determined rules.  We first assume that the number of bits is 

\begin{eqnarray}
\label{nfle}
N = \frac{A}{l^2},
\end{eqnarray}
where  $A$ is the area of the holographic screen ($A=4\pi r^2$) and $l$ is the fundamental length. Thus, considering a particle (or a spherical mass distribution) with mass $M$ and energy, $E = M c^2$,  placed in the center of the  holographic screen and combining  Eqs.(\ref{un}), (\ref{ge}) and (\ref{nfle}), we find

\begin{eqnarray}
a &=& \frac{5 - 3 q}{2} \, \frac{l^2 c^3}{\hbar} \;\; \frac{ M}{r^2}\nonumber\\ \nonumber\\
&=& G \, \frac{ M}{r^2},
\end{eqnarray}
with the Newton constant $G$ being recognized as

\begin{eqnarray}
\label{gg}
G = \frac{5 - 3 q}{2} \, \frac{l^2 c^3}{\hbar}.
\end{eqnarray}
We would like to remark here that in Eq.(\ref{nfle}) we have considered the fundamental length $l$ as the space time unit instead of the Planck length $l_p$. 
Using Eqs.(\ref{gc}) and (\ref{gg}) we can establish a relation for the square of the Planck length as

\begin{eqnarray}
\label{lpq}
l_p^2 = \frac{5 - 3 q}{2} \, l^2.
\end{eqnarray}
Comparing  Eq.(\ref{lpq}) with  (\ref{lp}), we can identify the positive numerical factor $f$  as

\begin{eqnarray}
\label{f}
f = \frac{5 - 3 q}{2}.
\end{eqnarray}
From Eq.(\ref{f}), we can note that the value of the positive numerical factor $f$ is controlled by the nonextensive parameter $q$ of the bits on the holographic screen. For example, for $q=1$ we have $ f=1$ and, consequently, $l=l_p$. For $q=5/3$ we have $f=0$ and, consequently, $l$ diverges ($l  \rightarrow \infty)$.   It is important to stress that although $f$ varies with the nonextensive parameter 
$q$, Eq.(16), the fundamental length $l$ also 
varies with $q$ in the sense that the product, Eq.(2), remains constant
(as we said in the last part of Section 2, the Planck length $l_p$ is a fixed value). Therefore, 
the fact that $ f$ may depend on $q$  does not 
contradict  the  hypothesis that Eq.(2) is a universal relation as mentioned in the introduction.

This last result allows us to imagine a curious hypothetical case that is when a nonextensive physical system has the value of the parameter $q$ approaching to $5/3$ and consequently the fundamental length  becoming large comparable to the usual values of our real world. In this way, as the fundamental length, in its proper meaning, drives the quantum structure of space time, then we hope that this hypothetical system should exhibit quantum space time effects in a macroscopic domain or in a low energy regime. This is an amazing result and the search of such possible gravitational systems can be an important task.

It is important to mention here that Cantcheff and Nogales\cite{CN} have written the Tsallis' entropy in the form

\begin{eqnarray}
\label{tbh}
S_q = k_B \frac{\Gamma^{q-1} - 1}{q-1},
\end{eqnarray}
where $\Gamma$ is proportional to the volume, $\Gamma\approx V$. Then for $q=\frac{5}{3}$ the Tsallis' entropy $(\ref{tbh})$ scales as the area which is just the expression for the black hole area entropy law. Consequently, due to the nonextensivity of entropy (\ref{tbh}), we have the emergence of a holographic screen which is a fundamental component in the Verlinde's formalism. Also, according to the hypothetical quantum gravity, black holes must exhibit macroscopic quantum phenomena. Therefore, the equality between the value of $q$ for the Tsallis'entropy in the area scaling behavior and our result in the Verlinde's formalism for macroscopic quantum regime indicates that bits may follow a nonextensive statistical mechanics. A detailed discussion about nonextensivity and entropy area law can also be found in \cite{FT}.

Bringing  Eq.(\ref{lpq}) in the form

\begin{eqnarray}
\label{lfq}
l = l_p \,\sqrt{\frac{2}{5 - 3 q}},
\end{eqnarray}
we then plot in figure 1 the ratio $l/l_p$ as a function of the nonextensive parameter $q$. We can observe that for $ q < 1$ we have $ l < l_p$  and for $ q > 1$ we have $ l > l_p$. 

It is interesting to point out the resemblance between Eq.(\ref{lfq}) and the formula for the nonextensive critical wavelength, $\lambda_c \,$, in the Jeans' criterion for gravitational instability in the self-gravitating systems. In Eq.(\ref{lfq}), the Planck length, $l_p \,$, corresponds to the Jeans' length, $\lambda_j \,$, and the fundamental length, $l \,$, corresponds to the critical wavelength, $\lambda_c \,$. For more detail, see, for example, reference \cite{Jeans}.

By using the holographic principle, i.e. the nonextensivity of the system being mapped on the bits of the holographic screen, we can assume that, at first, a particular gravitational system and the corresponding bits on the holographic screen both have the same nonextensive parameter $q$. Therefore, gravitational systems or cosmological models can, in principle, help us in determining the $q$-parameter.  For example, from the cosmic background radiation data obtained by the Explorer Satellite, Tsallis, S\'a Barreto and Loh\cite{TBL} estimated a bound for the nonextensive $q$-parameter. They found an upper limit for $q$ as

\begin{eqnarray}
\label{bound}
\left|q-1 \right| < 3.6 \times 10^{-5}.
\end{eqnarray}
To calculate a value for $f\,$ we write Eq.(\ref{f}) in the form

\begin{eqnarray}
\label{fa}
f = 1 - \frac{3}{2} \, (q-1).
\end{eqnarray}
Using Eq.(\ref{fa}) we obtain, with the error determined by Eq.(\ref{bound}), a bound for $f$ as

\begin{eqnarray}
\label{nfl}
\left|f-1 \right| < 5.4 \times 10^{-5},
\end{eqnarray}
in which specifies a possible value for $f$ that is $ f \approx 1 $ and consequently $l\approx l_p$.  Here, it is interesting to comment that Tirnakli and Torres\cite{TT}, by using early universe test, found a bound for $q$ as $ \, \left|q-1 \right| < 4.01 \times 10^{-3}\,$,  which is certainly close to the result, Eq.(\ref{bound}).

\begin{figure}[t]
\renewcommand{\baselinestretch}{1.0}
\begin{center}
\includegraphics[totalheight=7cm]{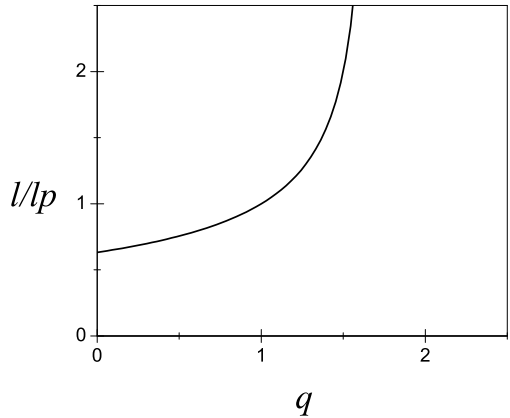}
\caption{\label{fig}The ratio $l/l_p$ is shown as a function of the nonextensive parameter $q$. For $ q=1$ we have $l=l_p$ and for $q=5/3$ the fundamental length diverges($l  \rightarrow \infty)$.}
\end{center}
\end{figure}

\section{Conclusions}

We have applied a nonextensive equipartition law of energy in the Verlinde's formalism. As an important result, we have associated the numerical positive factor $f$  with the nonextensive parameter $q$, Eq.(\ref{f}). In this way, the parameter $q$ that appears in the Tsallis' statistics would provide a connection between the Planck length with the fundamental length. In particular, we have shown that the $q$-dependence of the fundamental length, Eq.(\ref{lpq}), and the numerical positive factor, Eq.(\ref{f}), is, in principle, physically acceptable in the interval $ 0 \leq q < 5/3\,$ where the nonextensive equipartition theorem is valid.
We would like to point out that our approach can answer the question made by Klinkhamer about the possibility to design an experiment which  determines the value of the fundamental length. From our perspective,  we can say that the experiment already exists (see, for example, references \cite{TBL}, \cite{TT}, \cite{TBD}, \cite{PPV} and \cite{CT2} ) and consists of measuring the value of the nonextensive parameter $q$ from experimental data. 

It has been pointed out\cite{Jeans, gc} that the observed velocity distribution of galaxy clusters deviates from Maxwellian distribution and can be fitted well by the Tsallis' nonextensive distribution with $ q=0.23$.
 If the values of $q$ deviate from unity for different physical systems then our model predicts that the positive numerical factor and the fundamental length will depend on the degree of nonextensivity of the bits of a particular gravitational system. We can also mention that  stellar polytropes\cite{PP2} and black holes\cite{OD} are known examples of systems with $q$ possibly deviating from unity. By invoking the holographic principle, we can possibly avoid unphysical  results e.g. an isolated particle (a system without any interaction) with a finite mass and $q$ near $5/3$ consequently having the fundamental length of the space time close to infinity. Therefore,  it is important to observe that, as a consequence of our model, the properties of the quantum space time will also depend on the degree of nonextensivity of the bits of a particular gravitational system. As a perspective for future studies, it would be interesting to investigate the physical consequences of the adoption of a nonextensive statistical mechanics for the bits, not only in the fundamental length concept but, as an example, in the cosmological context.

\section{Acknowledgments}
The author would like to thank  Prof. F. Klinkhamer and  Prof. E. M. C. Abreu for useful discussions. The author would also like to thank Prof. U.Tirnakli for sending reference\cite{TT}.


\begin{thebibliography} {99}

\bibitem{Klink}
F. R. Klinkhamer,  Class. Quant. Grav. 28, 125003 (2011); F. R. Klinkhamer, JETP Lett. 86, 73 (2007); See also  A. E. Shalyt-Margolin,  
arXiv: gr-qc/1102.5084.

\bibitem{Sa}
H. Sahlmann, Class. Quant. Grav. 28, 015006 (2011).

\bibitem{SM}
L. Shao and  B. Q.  Ma, arXiv: hep-th/1006.3031.

\bibitem{Tsa}
C. Tsallis, J. Stat. Phys. 52, 479 (1988).

\bibitem{Ver}
E. Verlinde, JHEP 1104, 029 (2011); 

\bibitem{Levy}
P. A. Alemany and D. H. Zanette, Phys. Rev. E 49, 956 (1994); P. A. Alemany and D. H. Zanette, Phys. Rev. Lett. 75, 366 (1995).

\bibitem{CT}
C. Anteneodo and C. Tsallis, J. Mol. Liq. 71, 255 (1997).

\bibitem{Ng}
 R. Silva and J. S. Alcaniz, Phys. Lett. A 313, 393 (2003);  H. P. de Oliveira and I. Dami\~ao Soares, Int. Jour. Mod. Phys. D, 17 nos.3 \& 4, 541 (2008); S. H. Hansen, D. Egli, L. Hollenstein and C. Salzmann, New Astronomy 10, 379 (2005); R. Silva and J. S. Alcaniz, Physica A 341, 208 (2004); C. Tsallis, Chaos, Soliton and Fractals 13, 371 (2002).

\bibitem{sev}
For example we can cite: S. H. Hendi and A. Sheykhi, Phys. Rev. D 83, 084012 (2011); J. Ananias Neto, arXiv: hep-th/1009.4944; D. Momeni,  arXiv: physics.gen-ph/1009.2182; C. Bastos, O. Bertolami, N. C. Dias and J. N. Prata, Class. Quant. Grav. 28, 125007 (2011);  P. Nicolini, Phys. Rev. D 82, 044030 (2010); Y. Tian and X. N. Wu, Phys. Rev. D 83, 021501(R) (2011);  X. Li and Z. Chang, Commun. Theor. Phys. 55, 733 (2011); Y. Tian and X. N. Wu, Phys. Rev. D 81, 104013 (2010); C. Gao, Phys. Rev. D 81, 087306 (2010); R. G. Cai, L. M. Cao and N. Ohta, Phys. Rev. D 81, 061501(R) (2010); L. Modesto and A. Randono, arXiv: hep-th/1003.1998;  M. R. Setare and D. Momeni, arXiv: physics.gen-ph/1004.0589; Y. Zhao,  arXiv: hep-th/1002.4039; I. V. Vancea and M. A. Santos, arXiv: hep-th/1002.2454.

\bibitem{Ko}
A. Kobakhidze, Phys. Rev. D 83, 021502 (2011); A Kobakhidze, hep-th/1108.4161.

\bibitem{Ch}
M. Chaichian, M. Oksanen and A. Tureanu,  Phys.Lett B 702, 419 (2011); M. Chaichian, M. Oksanen and A. Tureanu,  hep-th/1109.2794. 

\bibitem{Pa}
T. Padmanabhan, Mod. Phys. Lett. A, vol. 25, no. 14, 1129 (2010) and references therein.

\bibitem{Unr}
W. G. Unruh, Phys. Rev. D 14, 870 (1976).

\bibitem{Codata}
P. J. Mohr, B. N. Taylor and D. B. Newell, Rev. Mod. Phys. 80, 633 (2008).

\bibitem{PL}
A. R. Plastino and J. A. S. Lima, Phys. Lett. A 260, 46 (1999).

\bibitem{SPL}
R. Silva, A. R. Plastino and J. A. S. Lima, Phys. Lett. A 249, 46 (1998); J. A. S. Lima, R. Silva and A. R. Plastino, 
Phys. Rev. Lett. 86, 2938 (2001).

\bibitem{MPP}
S. Martinez, F. Pennini and A. Plastino, arXiv: cond-mat/0006139.

\bibitem{CN}
M. B. Cantcheff and J. A. C. Nogales, Int. Jour. Mod .Phys. A, 21 no.15, 3127 (2006).

\bibitem{FT}
F. Caruso and C. Tsallis, Phys. Rev. E 78, 021102 (2008).

\bibitem{Jeans}
D. Jiulin, Phys. Lett. A 320, 347 (2004).

\bibitem{TBL}
C. Tsallis, F. C. S\'a Barreto and E. D. Loh, Phys. Rev. E 52, 1447 (1995). 

\bibitem{TT}
U. Tirnakli and D. F. Torres, Physica A 268, 225  (1999).

\bibitem{TBD}
U. Tirnakli, F. Buyukkilic and D. Demirhan, Phys. Lett. A 245, 62 (1998).

\bibitem{PPV}
A. R. Plastino, A. Plastino and H. Vucetich,  Phys. Lett. A 207, 42 (1995).

\bibitem{CT2}
C. Tsallis, Braz. J. Phys. vol.29, no.1, 1 (1999). 

\bibitem{gc}
A. Lavagno, G. Kaniadakis, M. Rego-Monteiro, P. Quarati and C. Tsallis, Astrophys. Lett. Commun. 35, 449 (1998).

\bibitem{PP2}
A. Plastino and A. R. Plastino, Phys. Lett. A 174, 384 (1993).

\bibitem{OD}
H. P. de Oliveira and I. Dami\~ao, Phys. Rev. D 71, 124034 (2005).


\end{thebibliography}
\end{document}